\documentclass[lettersize,journal]{IEEEtran}
\usepackage{amsmath,amsfonts}
\usepackage{algorithmic}
\usepackage{algorithm}
\usepackage{array}
\usepackage[caption=false,font=normalsize,labelfont=sf,textfont=sf]{subfig}
\usepackage{textcomp}
\usepackage{stfloats}
\usepackage{url}
\usepackage{verbatim}
\usepackage{graphicx}
\usepackage{cite}

\usepackage[wby]{callouts}
\usepackage{url}
\usepackage[hidelinks]{hyperref}
\usepackage{overpic}
\usepackage{siunitx}
\usepackage{pict2e} 

\begin{document}
\title{Quasiparticle Generation-Recombination Noise in the Limit of Low Detector Volume}
\author{%
J. Li, P. S. Barry, T. Cecil, C. L. Chang, K. Dibert, R. Gualtieri, M. Lisovenko, Z. Pan, V. Yefremenko, G. Wang, and J. Zhang

\thanks{
(\textit{Corresponding author: Juliang Li)}}
\thanks{J. Li is with the Argonne National Laboratory, 9700 South Cass Avenue, Lemont, IL, 60439, USA (email: juliang.li@anl.gov)}%
\thanks{P. Barry is with Cardiff University, Cardiff CF10 3AT, UK (email: barryp2@cardiff.ac.uk)}
\thanks{T. Cecil; R. Gualtieri, M. Lisovenko, Z. Pan, V. Yefremenko, G. Wang and J. Zhang are with the Argonne National Laboratory, 9700 South Cass Avenue, Lemont, IL, 60439, USA (emails: cecil@anl.gov; rgualtieri@anl.gov; mlisovenko@anl.gov;  panz@anl.gov; yefremenko@anl.gov; gwang@anl.gov; jianjie-zhang@outlook.com;  )}%
\thanks{C. L. Chang is with the Argonne National Laboratory, Argonne, IL 60439 USA, the University of Chicago, 5640 South Ellis Avenue, Chicago, IL, 60637, USA, and Kavli Institute for Cosmological Physics, U. Chicago, 5640 South Ellis Avenue, Chicago, IL, 60637, USA (email: clchang@kicp.uchicago.edu) }%
\thanks{K. Dibert is with the University of Chicago, 5640 South Ellis Avenue, IL, 60637, USA (email: krdibert@uchicago.edu) }%

}

\maketitle


\begin{abstract}
We have measured the quasiparticle generation-recombination (GR) noise in aluminium lumped element kinetic inductors with a wide range of detector volumes at various temperatures. The basic detector consists of meandering inductor and interdigitated capacitor fingers. The inductor volume is varied from 2 to 153~\SI{}{\micro\mathbf{\meter^{3}}} by changing the inductor width and length to maintain a constant inductance. We started with measuring the power spectrum density (PSD) of the detectors frequency noise which is a function of GR noise and we clearly observed the spectrum roll off at 10~kHz which corresponds to the quasiparticle lifetime. Using data from a temperature sweep of the resonator frequency we convert the frequency fluctuation to quasiparticle fluctuation and observe its strong dependence on detector volume: detectors with smaller volume display less quasiparticle noise amplitude. Meanwhile we observe a saturated quasiparticle density at low temperature from all detectors as the quasiparticle life time $\boldsymbol{\tau_{qp}}$ approaches a constant value at low temperature. 
\end{abstract}



\section{Introduction}
\begin{figure}[h!]
     \begin{overpic}[abs,unit=1pt,scale=.06,width=0.48\textwidth]{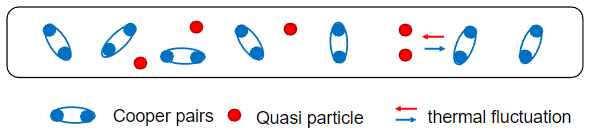}
     \put(0,5){\color{black}\large{(a)}}
     \put(-5,-125){\linethickness{0.5mm}\color{white}%
     \frame{\includegraphics[width=3.5in]{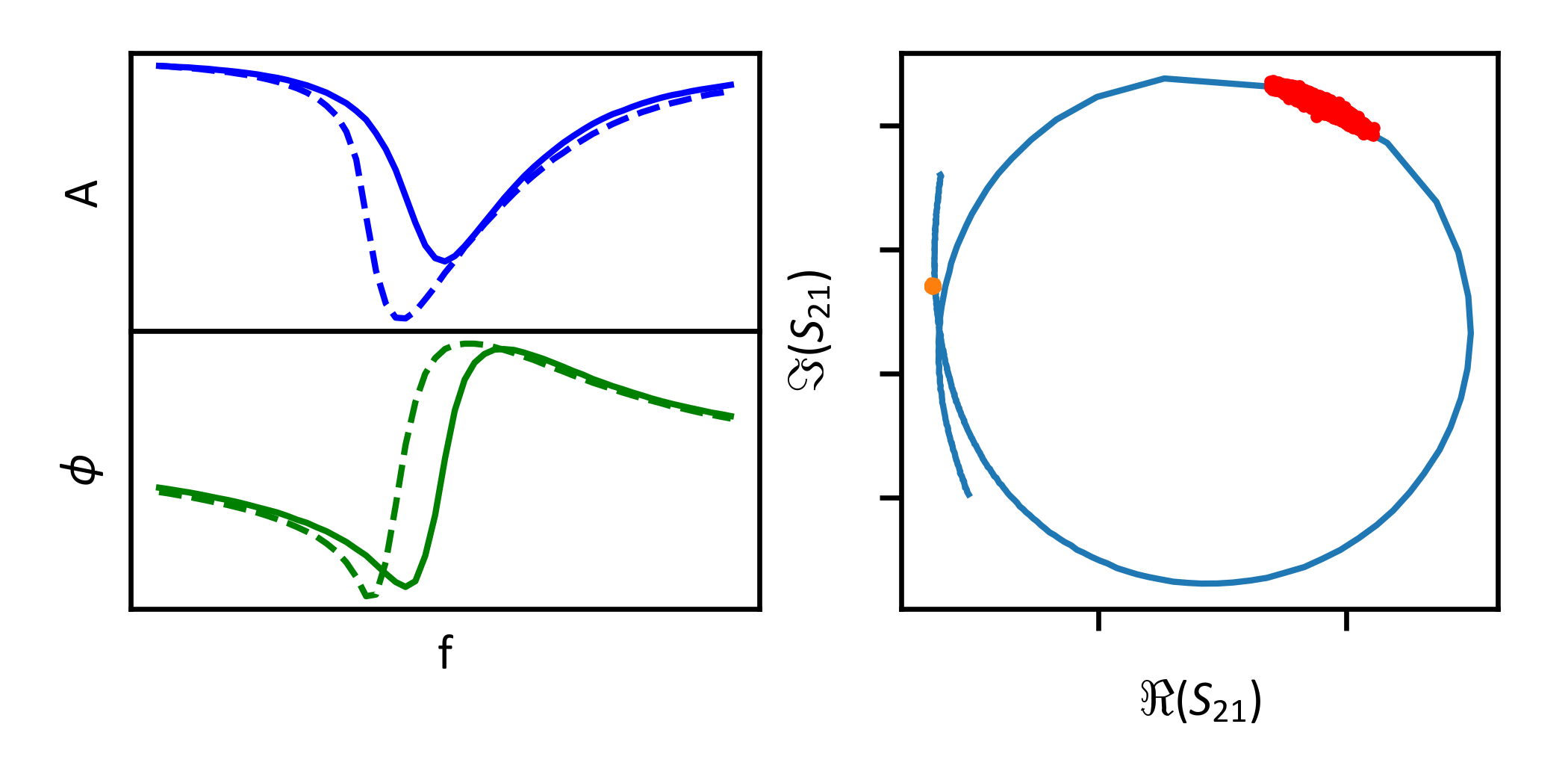}}}
     \put(20,-95){\color{black}\large{(b)}}
     \put(216,-95){\color{black}\large{(c)}}
     \end{overpic}\
     \vskip 110pt
     \caption{\footnotesize{\textbf{a}, schematic of a superconductor in thermal equilibrium which is a process of generation (red arrow) and recombination (blue arrow) of quasiparticles. \textbf{b}, exaggerated amplitude (blue) and phase noise (green) induced by the resonant frequency shift from the quasiparticle number change. These correspond to the noise in the radial and tangent direction of the resonance circle in (c). Solid lines stand for the state before quasiparticle noise and the dash lines stand for the change as result of the noise. \textbf{c}, example of a noise measurement with red dots showing 10,000 samples of the resonance frequency at the highest response point of the resonance circle. The oval shape of the noise envelope indicates that the phase noise is larger than amplitude noise\cite{thesisGao}.}}
\label{fig::noise}
\end{figure}

With over a decade of development effort, arrays of Microwave kinetic inductance detector (mKID) are found in a broad range of applications that require high-fidelity measurement of low energy signals\cite{mkid2003, thesisMazin,BASELMANS2007708,barends2007, mirzaei2020, barry2022, endo2012, shirokoff2012, wilson2020, brien2018, calvo2016}. The fundamental limit to the mKID detector’s sensitivity is the quasiparticle generation-recombination (GR) noise, which originates from the stochastic fluctuations in quasiparticle density as shown in the cartoon in Fig.~\ref{fig::noise} (a). The measured GR noise level depends on the detector volume, and it has been observed that any residual quasiparticle density imposes a practical lower limit on the achievable detector sensitivity, and is particularly important for detectors operating under ultra-low levels of optical loading. With an array of resonators designed to probe these effects directly, we investigate the GR noise level by varying detector volume over a wide range. Our result will suggest strategies to control and mitigate GR noise in future large-format arrays of mKIDs.

\begin{figure}[h!]
     \begin{overpic}[abs,unit=1pt,scale=.1,width=0.48\textwidth]{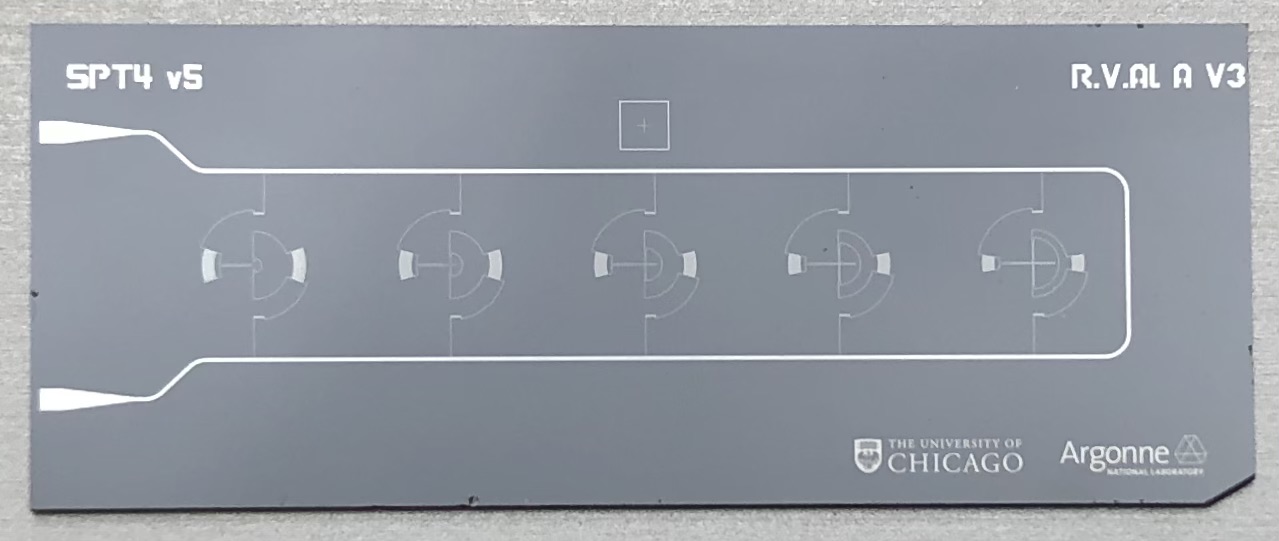}
     \put(11,14){\color{white}\large{(a)}}
     \put(60,-140){\color{black}%
     \frame{\includegraphics[scale=.1]{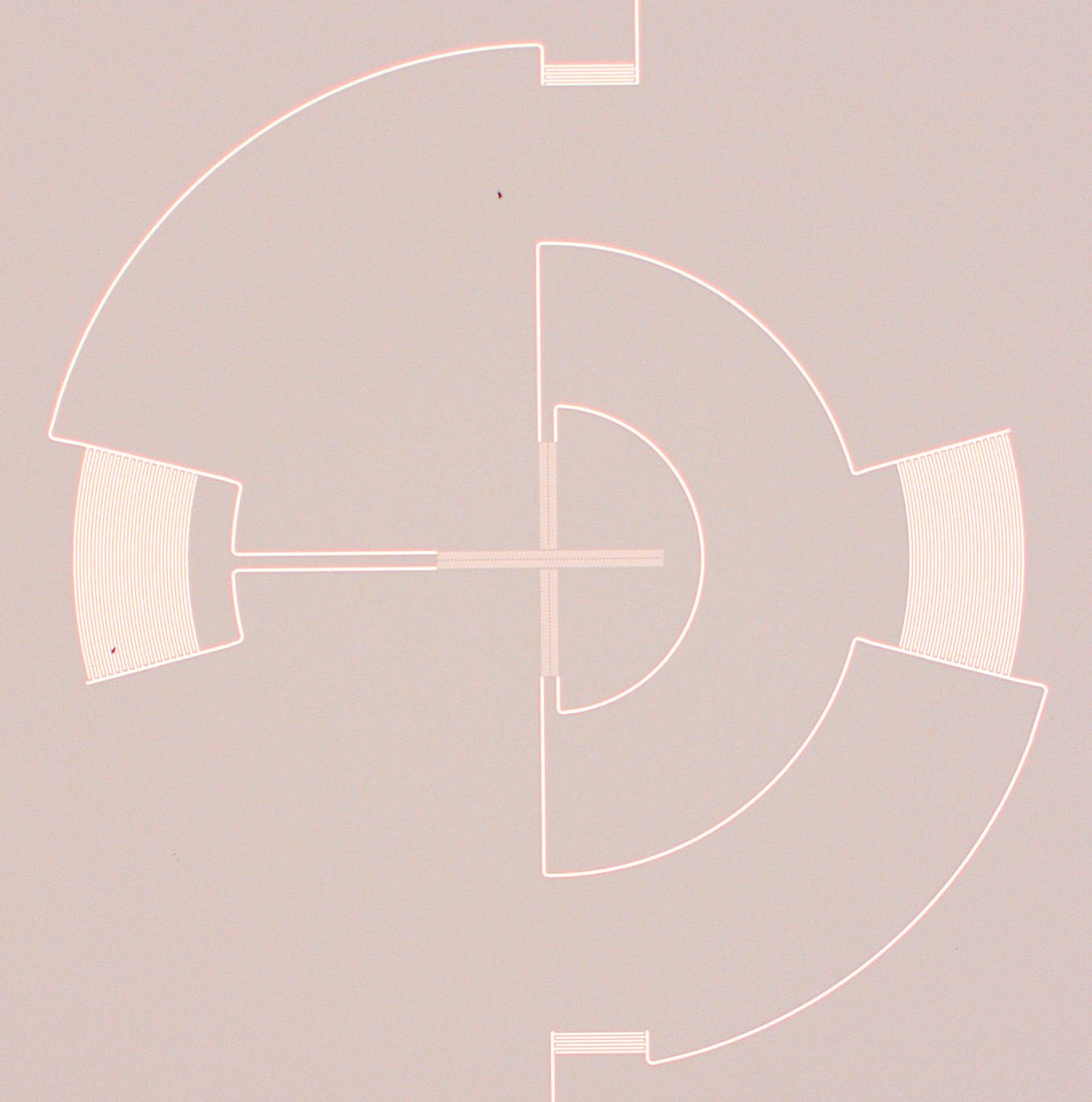}}}
     \put(5,-90){\color{black}%
     \frame{\includegraphics[scale=.08, angle = 90]{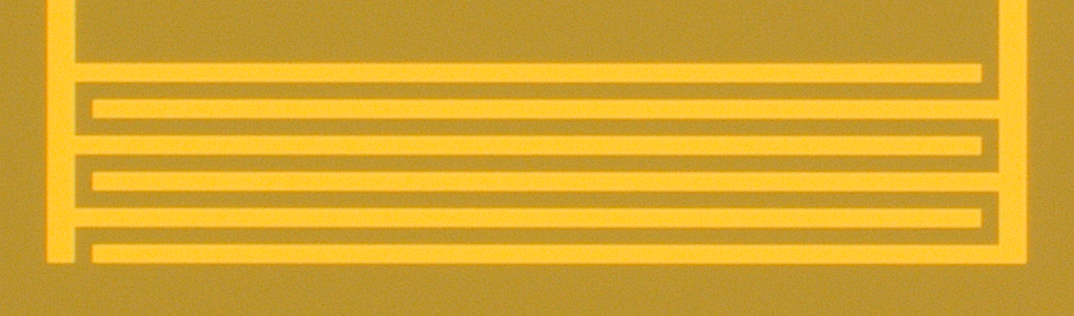}}}
     \put(180,-15){\color{white}\large{(b)}}
     \put(75,-50){\color{white}\large{$C_{x}$}}
     \put(170,-50){\color{white}\large{$C_{y}$}}
     \put(129,-50){\color{white}\small{Y-pol}} 
     \put(100,-82){\color{white}\small{X-pol}}
     \put(135,-200){\linethickness{0.5mm}\color{black}%
     \frame{\includegraphics[scale=.067]{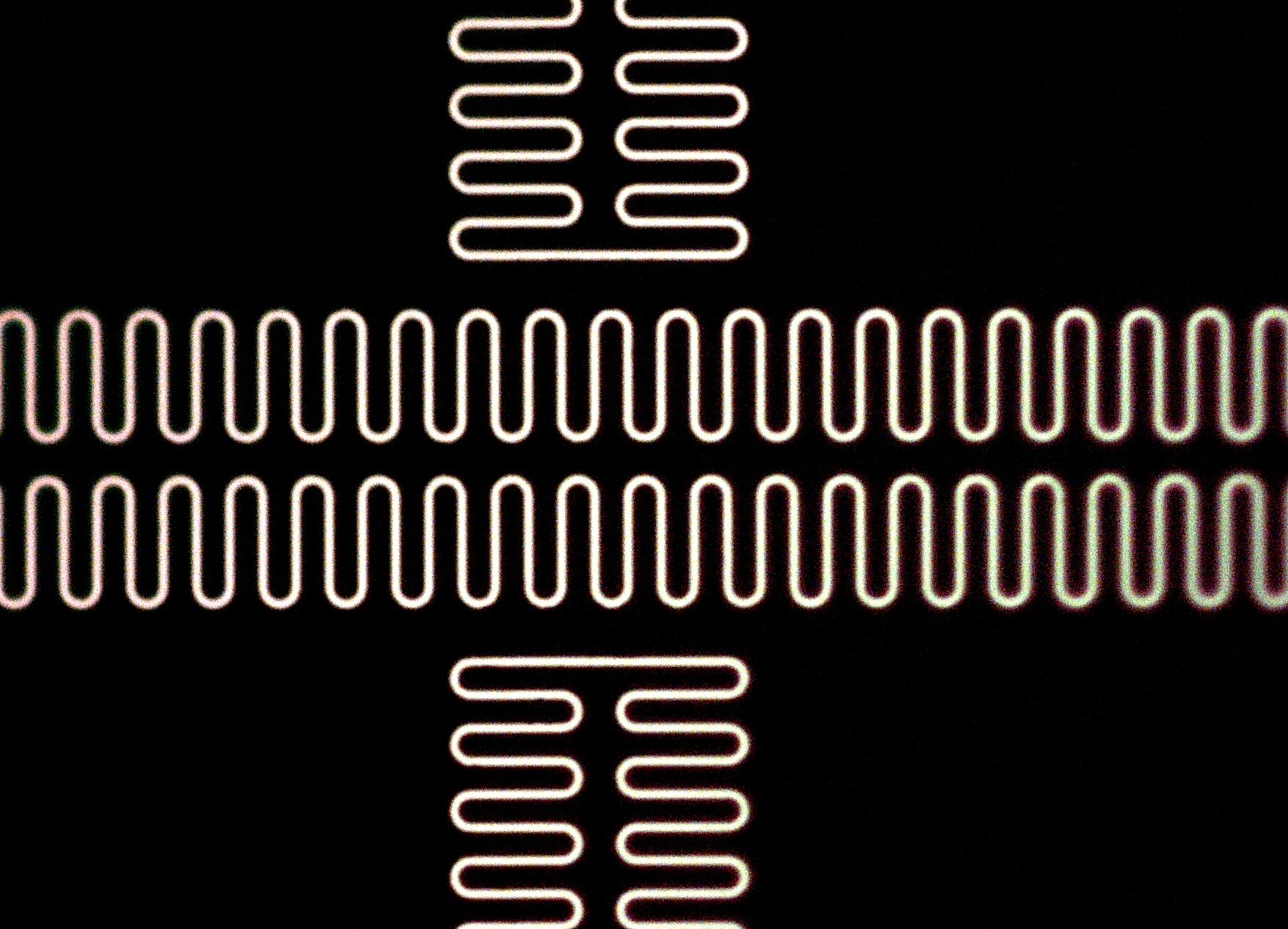}}}
     \put(5,-80){\color{white}\large{(e)}}
     \put(3,-200){\linethickness{0.5mm}\color{black}%
     \frame{\includegraphics[scale=.06]{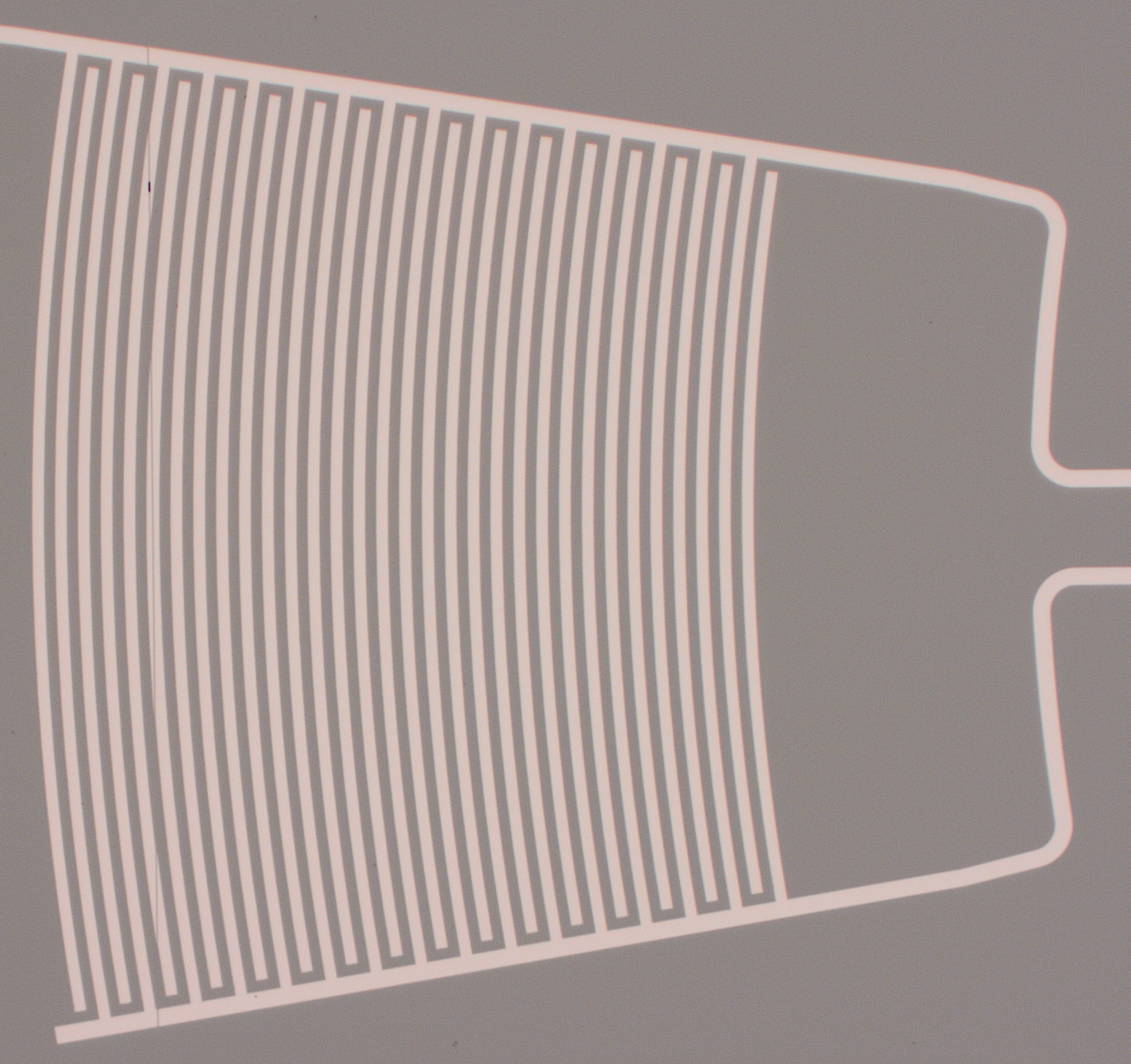}}}
     \put(145,-195){\color{white}\large{(d)}}
     \put(91,-195){\color{white}\large{(c)}}
      \linethickness{0.35mm}
     \multiput(75,-71)(0,-5){8}{\line(0,-1){1.8}}
     \multiput(125,-11)(-5,0){20}{\line(1,0){1.8}}
     \multiput(128,-72)(4,-4){12}{\line(1,-1){1.8}}
     \linethickness{0.35mm}
     \end{overpic}\
     \vskip 200pt
     \caption{\footnotesize{Images of the resonator array chip and components. \textbf{(a)} resoantor array which has five pixels and each pixel has two resonators named `X-pol' and `Y-pol' because of the orthogonal direction of the two inductors. \textbf{(b)} Zoom-in of one of the pixels. Left interdigital capacitor $C_{x}$ is connected to the horizontal (X) S-shaped inductor and the right interdigital capacitor $C_{y}$ is connected to the vertical (Y) S-shaped inductor. \textbf{(c)} Zoom in of an interdigital capacitor $C_{y}$. \textbf{(d)} Zoom-in of the S-shaped inductors. \textbf{(e)} the coupling capacitor that couples the detector to the readout line.}}
\label{fig::resodesign}
\end{figure}
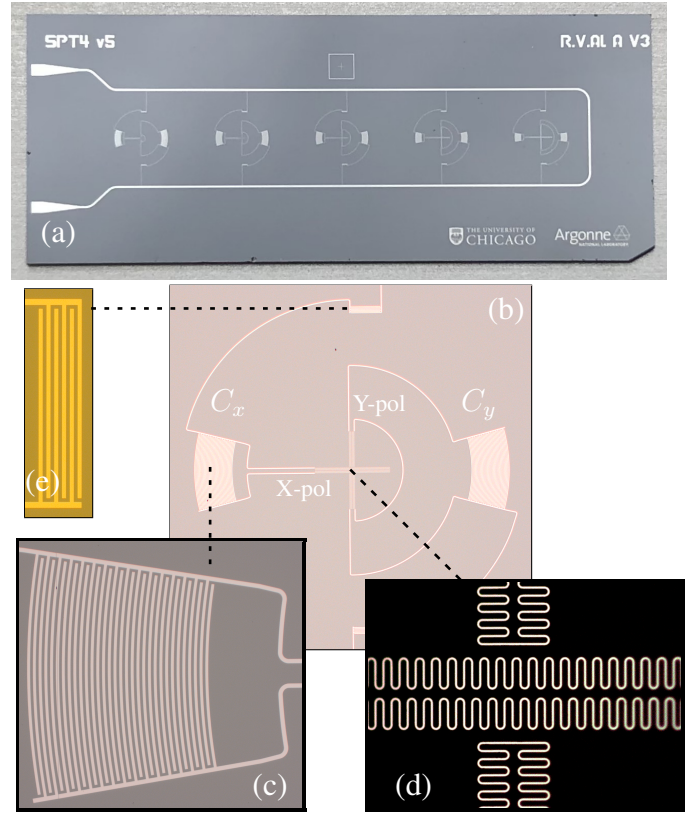

We start by measuring the power spectrum density of the frequency noise of an array of lumped LC superconducting resonators with stepped inductor volume. A resonator layout is shown in Fig.~\ref{fig::resodesign}. The resonator is coupled to the readout transmission line by means of a coupling interdigital capacitor which determines the resonator coupling quality factor $Q_{c}$. It is designed with low value so that the resonator ring down time $\tau_{r}$ is much shorter than the life time $\tau_{qp}$ of the quasiparticles in the resonator. Resonator frequencies are stepped by changing the resonator capacitance. Fig.~\ref{fig::noise}(b) shows the measured amplitude and phase of one resonator and change in response to quasiparticle change in the resonator. Noise from the quasiparticle fluctuation is measured with a standard homodyne detection setup and shown as red dots on the resonance circle in Fig.~\ref{fig::noise}(c).

\begin{figure}[h!]
     \begin{overpic}[abs,unit=1pt,scale=.1,width=0.48\textwidth]{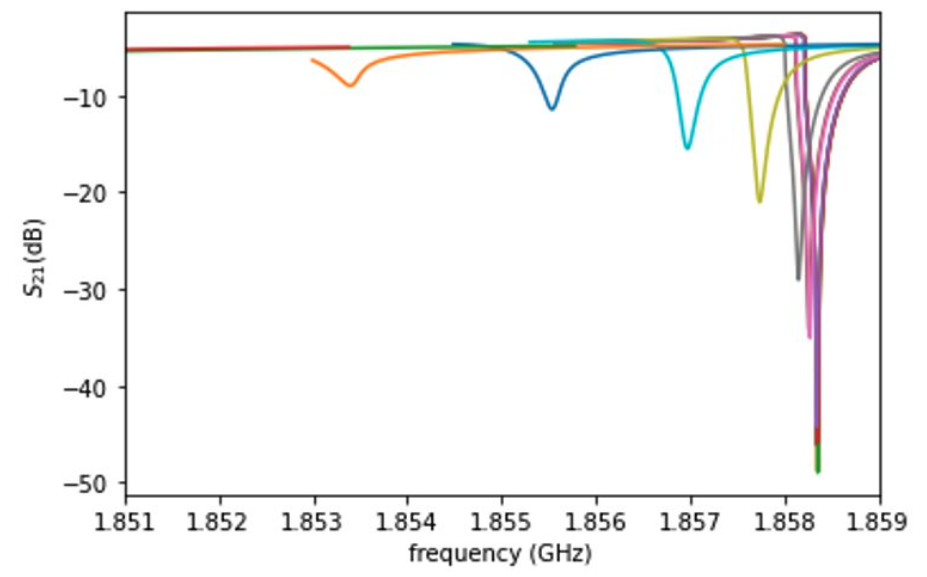}
     \put(36, 22){\color{white}%
     \frame{\includegraphics[scale=.25]{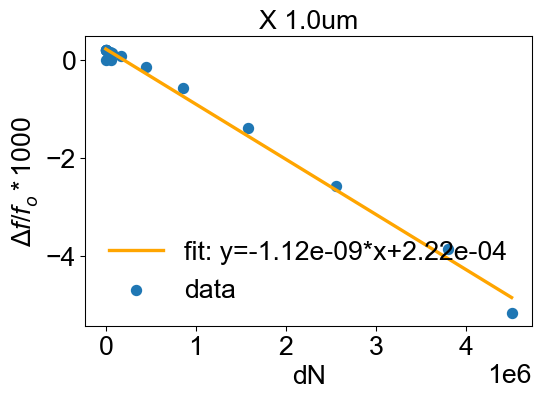}}}
     \end{overpic}\
     \vskip 3pt
     \caption{\footnotesize{Example of resonance dip change for temperatures between \SI{14} (green color) and \SI{270}{mK} (orange color). The resonance dip shifts to lower frequencies at higher temperatures. The inset shows $\Delta f/f_{o}=(f_{T}-f_{o})/f_{o}$ as a function of quasiparticle number (blue) along with a linear fit  (orange)}}
\label{fig::dfdNFit}
\end{figure}

The power spectral density of quasiparticle
fluctuations has a Lorentzian spectrum that is given by
\begin{equation}
S_{N}(\omega) = \frac{4N_{qp}\tau_{qp}}{1+(\omega\tau_{qp})^{2}}
\end{equation}
where $N_{qp}$ is the quasiparticle number in the detector and $\tau_{qp}$ is quasiparticle life time.
To determine $S_{N}(\omega)$ we measure the resonator frequency fluctuation $S_{f}(\omega)$ with a standard homodyne measurement setup as described in reference\cite{thesispete}. $S_{f}(\omega)$ is related to $S_{N}(\omega)$ by 
\begin{equation}
S_{f}(\omega) = S_{N}(\omega)\frac{(df/dN_{qp})^{2}}{1+(\omega\tau_{r})^{2}}
\label{eq::dfdn}
\end{equation}
where $df/dN_{qp} = df/(Vdn_{qp})$, $V$ is the detector volume and $n_{qp}$ is the quasiparticle density level. $\tau_{r}$ is the resonator ring down time give by $\tau_{r}=\frac{Q}{\pi f_{o}}$. The test resonator array resonances were measured at multiple temperatures (Fig.~\ref{fig::dfdNFit}) and $n_{qp}$ is calculated for each temperature according to
\begin{equation}
    n_{qp} = 2N_{o}\sqrt{2\pi k_{B}T\Delta}\text{exp}(-\Delta/k_{B}T)
\end{equation}
which is valid at $k_{B}T<\Delta$\cite{visserprl}. $N_{o}$ is the single spin density of states at the Fermi level ($1.72\times 10^{10}um^{-3}eV^{-1}$), $k_{B}$ Boltzmann's constant, T is the sample temperature and $\Delta$ is energy gap of the superconductor. Plotting the percentage frequency shift $\Delta f/f_{o}$ as a function of $n_{qp}$ shows a linear relationship and $df/dn_{qp}$ is calculated from the slope of the fitted liner regression. 

\section{detector array design}
\label{sec:design}
In our design the overall size of the chip is 2.5~cm$\times$1~cm (Fig.~\ref{fig::resodesign}) and we have five pixels with each pixel having a `X-pol' resonator and a `Y-pol' resonator. Within each pixel the `X-pol' (`Y-pol') resonators have same inductance while the `Y-pol' capacitor has slight larger capacitance than the one for `X-pol'. The inductor line width for each pixel is varied from \SI{0.2} to \SI{1.8} {\micro\meter} in steps of \SI{0.4} {\micro\meter} across the five pixels. The average capacitance among the two resonator are stepped and the pixel with an inductor line width of \SI{1.8} {\micro\meter} has the largest capacitance. With our designed configuration we expect five groups of resonances corresponding to the five pixels. Within each group we expect two resonances corresponding to `X-pol' and `Y-pol' with the `Y-pol' on the lower frequency side. The interdigital capacitor arrangement plays an important role in the resonator placing as it is the major parameter we use to set the resonators in the frequency domain. Details of the design parameter values can be found in table \ref{tab:pars}

\section{Fabrication}
The detector arrays are fabricated on a $2''$ high intrinsic resistivity silicon wafer. To achieve the required \SI{400}{nm} line width resolution, the resonators are patterned using e-beam lithography with a resolution of \SI{10}{nm}.

The fabrication process begins with vacuum baking the bare wafer at \SI{120}{\degreeCelsius} for three minutes to remove moisture. E-beam resist (PMMA 950 A4) is spun on the wafer at a speed of \SI{4000} {rpm} for 40 seconds and then baked at \SI{180} {\degreeCelsius} for three minutes afterward. The entire detector array is patterned in a single lithography layer with a dose of \SI{720} {\micro C/cm^2}. A ratio of 5:95 between optimal contrast and uniform clearing is applied in the proximity effect correction to maintain a uniform exposure across the array pattern. After patterning the wafer is developed with MIBK and IPA with ratio of 1:3 for 40 seconds, rinsed in IPA for 10 seconds and then rinsed in DI water. Following development a 20 seconds oxygen descum process is applied to remove leftover e-beam resist. A \SI{30}{nm} thick Al film is then deposited at a rate of \SI{0.17}{nm/second} via DC magnetron sputtering in Ar at a process pressure of  \SI{3}{mTorr}. The sputtering system has a base pressure of \SI{1.7E-8} {Torr} and he applied cathode voltage is \SI{300}{volt}. The final processing step is an overnight soak in 1165 Remover for lift-off.

\section{test setup}
\begin{figure}[h!]
     \begin{overpic}[abs,unit=1pt,scale=.06,width=0.4\textwidth]{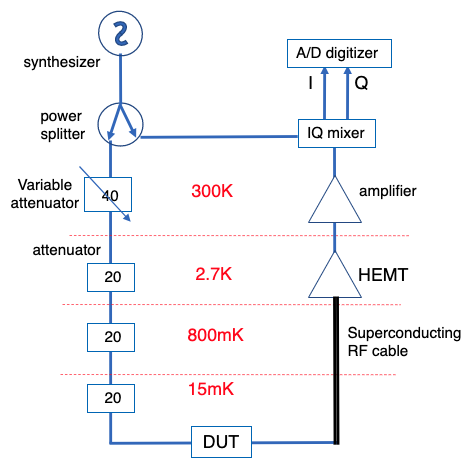}
     \end{overpic}\
     \caption{\footnotesize{A diagram of the homodyne readout system used for the noise measurement. There are 60dB of attenuation in total in the input line to reduce thermal noise coming from the higher temperature stages. Superconducting RF coaxial cable is used on the output line up to the HEMT amplifier stage to minimize resistive losses. The numbers in red color above the DUT mark the temperature of each stage. }}
\label{fig::meas}
\end{figure}

Noise measurements of the detector array are carried out with a homodyne measurement setup as illustrated in Fig.~\ref{fig::meas}. Each resonator is excited with a microwave tone generated with a synthesizer around the resonance frequency. The output signal is amplified with a cryogenic high electron mobility transistor (HEMT) amplifier mounted on the \SI{2.7}{K} stage followed by a room-temperature amplifier. The signal is then compared to the original signal using an IQ mixer. The output voltages I and Q of the IQ mixer carry the in-phase and quadrature amplitudes of the transmitted signal, respectively. When the carrier frequency is swept around resonant frequency of the resonators, the output I and Q trace out a resonance circle as shown in Fig.~\ref{fig::noise}\textbf{(c)}. For noise measurements the signal tone is fixed to be at a frequency just lower than the resonance frequency which has the highest noise fluctuation, and the fluctuations $\delta$I(t) and $\delta$Q(t) are digitized over a \SI{0.5}{s} interval using a sample rate $F_{s}$ = \SI{200}{kHz} for high frequency noise and a \SI{10}{s} interval using a sampling rate of $F_{s}$ = \SI{2}{kHz} for low frequency noise.

\section{test result}
Fig.~\ref{fig::resodark} shows the tested resonator distribution from the second version of the resonator design. The first challenge in this research project was matching the resonators with their line widths. Result from the first version of the design were very far from the simulation results and we could not identify the resonators with enough confidence. However, these results provided enough feedback to adjust the resonator capacitance to get the right distribution in the second version. Parameters for the detector array are calculated from the measurement data and listed in Table \ref{tab:pars}: Width is the inductor line width for each pixel, Tin stands for the length of the interdigital capacitor finger length, Vol is the inductor volume, $\tau_{r}$ is the resonator ring down time, $Q_{x}$ ($Q_{y}$) is the resonator quality factor for `X-Pol' (`Y-pol') detector. $Q_{i, x}$ ($Q_{i, y}$) is the intrinsic quality factor for `X-Pol' (`Y-pol') detector and $Q_{c, x}$ ($Q_{c, y}$) is the coupling or external quality factor. $f_{o}$ is the resonant frequency of the resonator, and $\frac{\partial f}{\partial N_{qp}}$ is the frequency shift with respect of quasiparticle number.
\begin{figure}[h!]
     \begin{overpic}[abs,unit=1pt,scale=.1,width=0.48\textwidth]{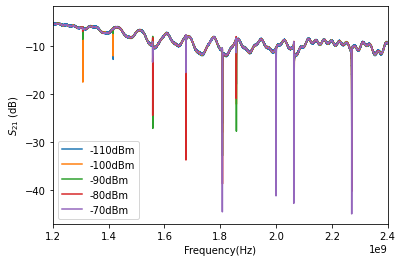}
     \put(41,100){\color{black}\small{0.2 Y}}
     \put(62,115){\color{black}\small{0.2 X}}
     \put(85,75){\color{black}\small{0.6 Y}}
     \put(105,55){\color{black}\small{0.6 X}}
     \put(125,40){\color{black}\small{1.0 Y}}
     \put(139,70){\color{black}\small{1.0 X}}
     \put(157,40){\color{black}\small{1.4 Y}}
     \put(175,30){\color{black}\small{1.4 X}}
     \put(205,30){\color{black}\small{1.8 Y}}
     \end{overpic}\
     \caption{\footnotesize{Amplitude of a probe tone frequency sweep showing the frequency distribution of the reduced volume detector array. The resonator distribution matches the designed distribution set by their respective capacitor capacitance. The missing 1.8~X resonator is likely due to a fabrication defect. The 1.8~Y resonator is not used in the data analysis to keep the symmetry of the result between the 'X-pos' and 'Y-pos'. Identities of each resonator are indicated next to their resonance dips with the number indicating their inductor line width and the letter for their polarization. Numbers in the legends stands for VNA output power during measurement.}}
\label{fig::resodark}
\end{figure}
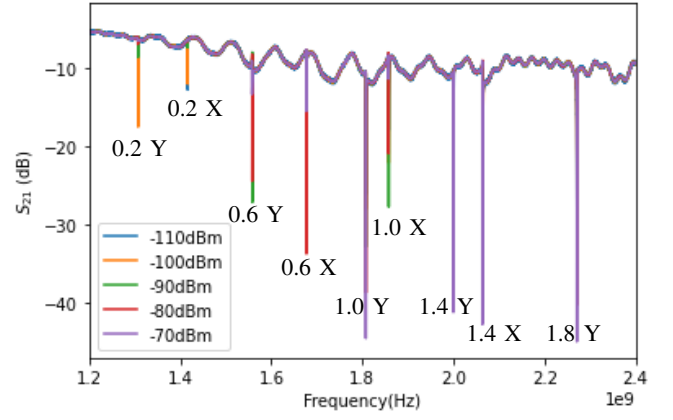
\begin{table}[!t]
\caption{Parameters of the eight resonators discussed in this paper. All parameters are given for a temperature of \SI{15}{mK}\label{tab:pars}. }
\centering
\begin{tabular}{|c||c|c|c|c|c|}
\hline
Width [um] & 0.2 & 0.6 & 1.0 & 1.4 & 1.8\\
\hline
$\text{Tin}_{X}$ & 620 & 540 & 460 & 380 & 300\\
\hline
$\text{Tin}_{Y}$ & 640 & 560 & 480 & 400 & 320\\
\hline
Vol [um$^{3}$] & 1.92 & 16.65 & 45.84 & 95.85 & 153.06\\
\hline
$\tau_{r,X}$ [usec] & 2.36 & 0.94 & 0.92 & 0.58 & N/A\\
\hline
$\tau_{r,Y}$ [usec] & 2.31 & 1.79 & 0.66 & 0.51 & 0.29\\
\hline
$Q_{x}$ & 10512 & 4971 & 5354 & 3765 & N/A\\
\hline
$Q_{y}$ & 9528 & 8753 & 3770 & 3213 & 2034\\
\hline
$Q_{i,x}$ & 27282 & 49997 & 48712 & 39338 & N/A\\
\hline
$Q_{i,y}$ & 28783 & 54808 & 50233 & 42284 & 35641\\
\hline
$Q_{c,x}$ & 17102 & 5520 & 6016 & 4163 & N/A\\
\hline
$Q_{c,y}$ & 14244 & 10417 & 4076 & 3478 & 2157\\
\hline
$f_{o, X}$ [GHz] & 1.42 & 1.68 & 1.86 & 2.06 & N/A\\
\hline
$f_{o, Y}$ [GHz] & 1.31 & 1.56 & 1.81 & 2.00 & 2.27\\
\hline
$\frac{\partial f_{X}}{\partial N_{qp}}$ & -3.01e-8 & -2.54e-9 & -1.12e-9 & -5.20e-10 & N/A\\
\hline
$\frac{\partial f_{Y}}{\partial N_{qp}}$ & -2.61e-8 & -2.84e-9 & -8.63e-10 & -4.0e-10 & -1.2e-10\\
\hline
\end{tabular}
\end{table}

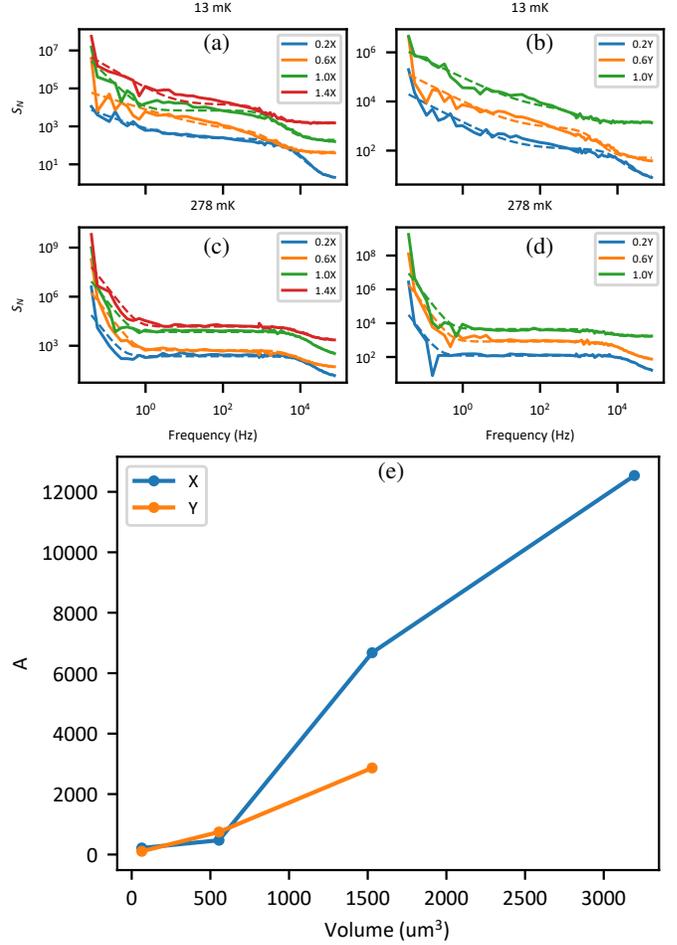
\begin{figure}[h!]
    \vskip 180pt
	\begin{overpic}[abs,unit=1pt,scale=.1,width=0.48\textwidth]{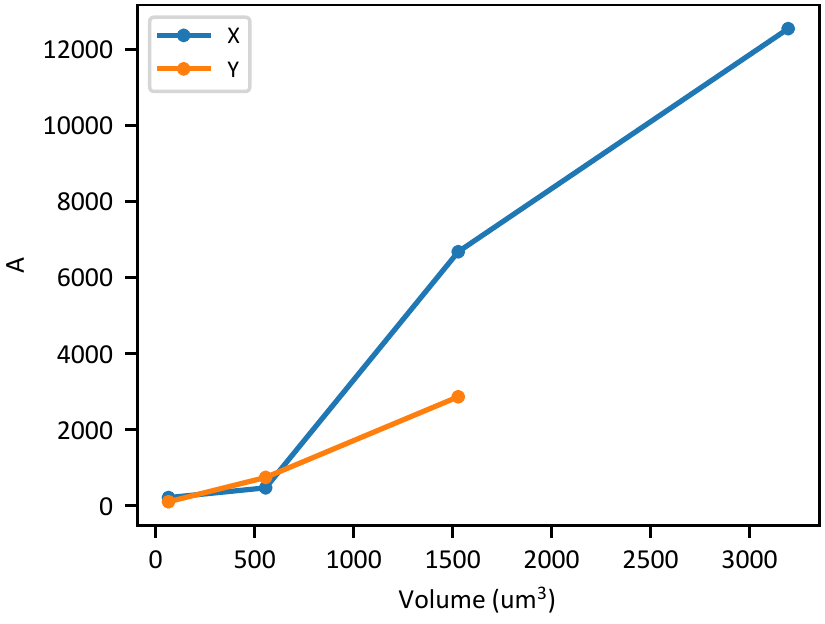}
		\put(120,265){\color{white}%
			\frame{\includegraphics[scale=1.1]{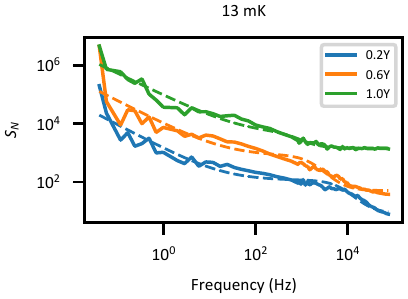}}}
		\put(0,265){\color{white}%
			\frame{\includegraphics[scale=1.1]{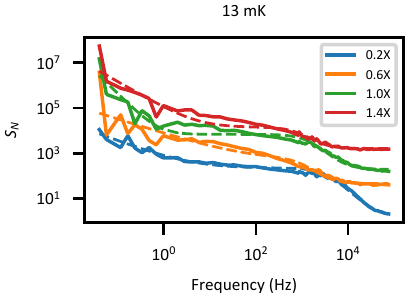}}}
		\put(120,190){\color{white}%
			\frame{\includegraphics[scale=1.1]{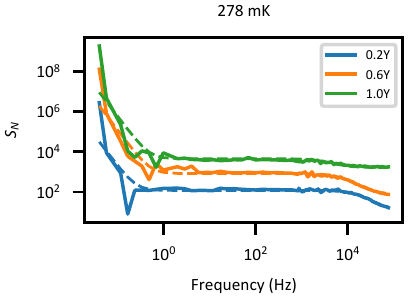}}}
		\put(0,190){\color{white}%
			\frame{\includegraphics[scale=1.1]{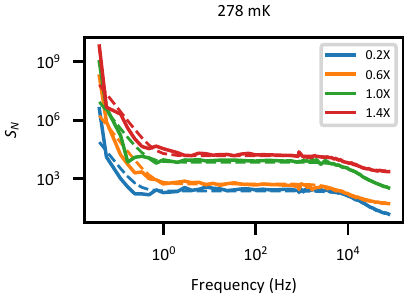}}}
		\put(74,340){\color{black}\small{(a)}}
		\put(196,340){\color{black}\small{(b)}}
		\put(74,263){\color{black}\small{(c)}}
		\put(196,263){\color{black}\small{(d)}}
		\put(140,178){\color{black}\small{(e)}}
	\end{overpic}\    
     \vskip -7pt
     \caption{\footnotesize{quasiparticle number noise power spectrum density $S_{N}$ at temperatures of \SI{15}{mK} for detectors of `X-pol' \textbf{(a)}, and for detectors of `Y-pol' \textbf{b}.  quasiparticle number noise power spectrum density $S_{N}$ at temperatures of \SI{278}{mK} for detectors of `X-pol' \textbf{(c)}, and for detectors of `Y-pol' \textbf{d}. The longer slope trend in the 15mK indicates the TLS noise which diminishes at temperatures above \SI{278}{mK}. The sharp slope on the low frequency range at T= \SI{278}{mK} indicates the 1/f noise which is a combination of the low frequency noise in our measurement electronics and the temperature fluctuation as the cooling power of our dilution fridge was fighting against the heating power we applied at the mixing chamber stage to keep the stage at \SI{260}{mK}. \textbf{(e)}, Amplitude of fitted $S_{N}$ as function of detector volume for both `X-pol' and `Y-pol' detectors at  temperature of \SI{15}{mK} and \SI{278}{mK}.}}
\label{fig::pswXY}
\end{figure}

The measured quasiparticle number noise $S_{N}$ for the two polarizations, `X-pol' and `Y-pol', are shown in Fig.~\ref{fig::pswXY}. At low temperature two-level system noise will cause fluctuations of the resonator frequency which shows as a long range slope on the PSD. At low frequency there is 1/f noise fluctuation from the measurement electronics and temperature fluctuations. All the PSD curves show the roll off (the knee) at \SI{10}{kHz} which is the quasiparticle generation-recombination frequency. Above the roll off the dominant noise source is the HEMT amplifier and the noise level should the same for all the resonators. However, our noise measurement only goes up to \SI{100}{kHz} which is determined by our digitizer speed. The noise floor will eventually line up once the sample frequency is high enough. The relative plateau level of the PSD on the vertical axis indicates the relative GR noise amplitude within each detector group and both X-pol and Y-pol groups show clear dependence of GR noise  amplitude on detector volume. As expected from equation~(\ref{eq::dfdn}) a smaller detector volume gives a lower noise amplitude and this matches with our measured PSD distribution. Sweep data for the detector with \SI{1.4} {\micro\meter} line width in the `Y-pol' group had a significant glitch that caused as error in the calculated PSD and we have removed it from the final result. Additionally, we don't see a clear roll off for all the `Y-pol' detectors. One possible reason is that `Y-pol' detector inductors are two pieces connected with a longer and wider arc; otherwise a bridge or cross over structure is required for the `Y-pol' inductor to pass over the `X-pol' inductor. This extra arc might have influenced the quasiparicle distribution and hence and the PSD. At \SI{278}{mK} `X-pol' $S_{N}$ curves have uneven spacing across detector volumes, which is absent from the `Y-pol' detectors at the same temperature.     

We applied a complete fitting of the PSD curves to equation~(\ref{eq::Stotal}) which includes the 1/f noise and the TLS noise to retrieve the quasiparticle noise amplitude and life time $\tau_{qp}$.
\begin{equation}
\label{eq::Stotal}
S_{xx}(f) = \left(\frac{A+Bf^{-n}}{1+(2\pi f\tau_{qp})^{2}}+C\right)
\end{equation}
Term $A$ stands for the quasiparticle noise amplitude and $Bf^{-n}$ stands for the sharp slope from the 1/f noise and the long slope from the TLS noise. The fitted line is plotted as dashed line on top of the PSD curve in Fig.~\ref{fig::pswXY}. The amplitude of $S_{N}$ is positively proportional to detector volumes for both groups. There is a sign of saturation of the noise amplitude at very low detector volume. Measurements with additional low detector volume is required to confirm this.

\begin{figure}[h!]
     \begin{overpic}[abs,unit=1pt,scale=.1,width=0.48\textwidth]{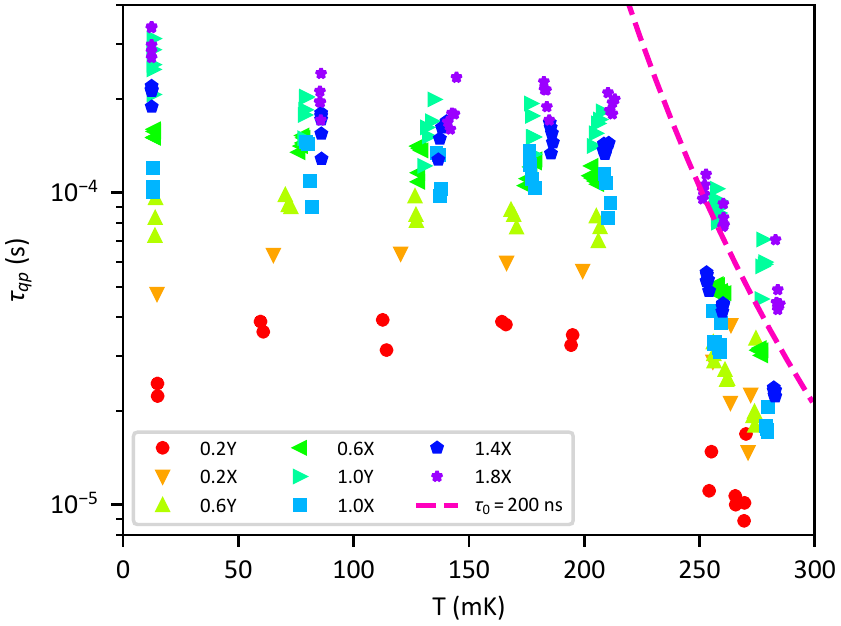}
     \end{overpic}\
     \caption{\footnotesize{Quasiparticle life time $\tau_{qp}$ as function of detector temperature for all detector volumes. At higher temperatures the qp lifetime of all detectors decreases inversely proportional to temperature as suggested by equation~(\ref{eq::tauT}). Due to the quasi particle population saturation at low temperature $\tau_{qp}$ flattens out instead of continuously going higher. The dashed purple line is an exponential fit of data above \SI{225}{mK}  to equation~(\ref{eq::tauT}) with $\tau_{o}=200$ns, which is an average life time value for all the detectors.}}
\label{fig::taut}
\end{figure}

Assuming a thermal distribution of quasiparticls and photons at low temperature, the average quasiparticle life time is given by\cite{kaplan1976}
\begin{eqnarray}
    \tau_{qp} &=& \frac{\tau_{o}}{\sqrt{\pi}}\left(\frac{k_{B}T_{c}}{2\Delta}\right)^{5/2}\sqrt{\frac{T_{c}}{T}}e^{\Delta/k_{B}T} \\
    &=& \frac{\tau_{o}}{n_{qp}}\frac{N_{o}(k_{B}T_{c})^{3}}{2\Delta^{2}}
    \label{eq::tauT}
\end{eqnarray}
where $T_{c}$ is the critical temperature of the superconductor and $\tau_{o}$ stands for the characteristic electron-phonon interaction time, which is material dependent. Equation~(\ref{eq::tauT}) predicts $\tau_{qp}$ increases exponentially with temperature as a result of reduced quasiparticle density at lower temperature. Fig.~\ref{fig::taut} plots the calculated quasiparticle life time $\tau_{qp}$ for each detector as a function of temperature. Each detector is readout with multiple powers which corresponds to multiple curves with the same marker. At higher temperatures all the curves show the exponential function of $\tau_{qp}$ with temperature and agree with theoretical predication. At low temperatures $\tau_{qp}$ shows signs of saturation which suggests residual quasiparticle density $n_{qp}$ in the detectors from nonequilibrium quasiparticle excitation\cite{catelani2011, martinis2009}. This is valid with all the detectors at all the readout powers we applied in the measurement and agrees with result in reference\cite{visserprl}. We also observed that for a given temperature $\tau_{qp}$ increases with larger detector volume. This is most likely related to the readout power difference across the detectors. Detectors tend to have higher responsivity with higher readout power and we have been using readout power below and close to the resonator bifurcation point for readout. Detectors with a very narrow inductor line width have a much lower bifurcation power so there is a broad readout power range across detectors.

\section{Conclusion}
Our test results have validated the theory that quasiparticle G-R noise strongly depends on detector volume and GR noise can be reduced by using smaller detector volume. For very low detector volumes the contribution from other noise sources will be significant and additional measure methods will be needed to distinguish the individual noise sources. We have been demonstrated that higher readout power could induce excess quasiparticle density rising\cite{visser2012} and a low noise first stage amplifier is necessary to reduce the readout power.  Also we have demonstrated saturation of the quasiparticle lifetime due to the residual quasiparticle density within each detector.
\section{acknowledgements}
\footnotesize{Work at Argonne, including use of the Center for Nanoscale Materials, an Office of Science user facility, was supported by the U.S. Department of Energy, Office of Science, Office of Basic Energy Sciences and Office of High Energy Physics, under Contract No. DE-AC02-06CH11357.}

\bibliographystyle{IEEEtran}

\bibliography{reff.bib}

\begin{thebibliography}{10}
\providecommand{\url}[1]{#1}
\csname url@samestyle\endcsname
\providecommand{\newblock}{\relax}
\providecommand{\bibinfo}[2]{#2}
\providecommand{\BIBentrySTDinterwordspacing}{\spaceskip=0pt\relax}
\providecommand{\BIBentryALTinterwordstretchfactor}{4}
\providecommand{\BIBentryALTinterwordspacing}{\spaceskip=\fontdimen2\font plus
\BIBentryALTinterwordstretchfactor\fontdimen3\font minus
  \fontdimen4\font\relax}
\providecommand{\BIBforeignlanguage}[2]{{%
\expandafter\ifx\csname l@#1\endcsname\relax
\typeout{** WARNING: IEEEtran.bst: No hyphenation pattern has been}%
\typeout{** loaded for the language `#1'. Using the pattern for}%
\typeout{** the default language instead.}%
\else
\language=\csname l@#1\endcsname
\fi
#2}}
\providecommand{\BIBdecl}{\relax}
\BIBdecl

\bibitem{mkid2003}
\BIBentryALTinterwordspacing
P.~K. Day, H.~G. LeDuc, B.~A. Mazin, A.~Vayonakis, and J.~Zmuidzinas, ``A
  broadband superconducting detector suitable for use in large arrays,''
  \emph{Nature}, vol. 425, pp. 817 EP --, Oct 2003. [Online]. Available:
  \url{http://dx.doi.org/10.1038/nature02037}
\BIBentrySTDinterwordspacing

\bibitem{thesisMazin}
B.~Mazin, ``Microwave kinetic inductance detectors,'' Ph.D. diss., Pasadena CA,
  2004.

\bibitem{BASELMANS2007708}
\BIBentryALTinterwordspacing
J.~Baselmans, S.~Yates, P.~{de Korte}, H.~Hoevers, R.~Barends, J.~Hovenier,
  J.~Gao, and T.~Klapwijk, ``Development of high-q superconducting resonators
  for use as kinetic inductance detectors,'' \emph{Advances in Space Research},
  vol.~40, no.~5, pp. 708--713, 2007. [Online]. Available:
  \url{https://www.sciencedirect.com/science/article/pii/S0273117707006692}
\BIBentrySTDinterwordspacing

\bibitem{barends2007}
R.~Barends, J.~J.~A. Baselmans, J.~N. Hovenier, J.~R. Gao, S.~J.~C. Yates,
  T.~M. Klapwijk, and H.~F.~C. Hoevers, ``Niobium and tantalum high q
  resonators for photon detectors,'' \emph{IEEE Transactions on Applied
  Superconductivity}, vol.~17, no.~2, pp. 263--266, 2007.

\bibitem{mirzaei2020}
\BIBentryALTinterwordspacing
M.~Mirzaei, E.~M. Barrentine, B.~T. Bulcha, and e.~a. Giuseppe~Cataldo,
  ``{µ-spec spectrometers for the EXCLAIM instrument},'' in \emph{Millimeter,
  Submillimeter, and Far-Infrared Detectors and Instrumentation for Astronomy
  X}, J.~Zmuidzinas and J.-R. Gao, Eds., vol. 11453, International Society for
  Optics and Photonics.\hskip 1em plus 0.5em minus 0.4em\relax SPIE, 2020, p.
  114530M. [Online]. Available: \url{https://doi.org/10.1117/12.2562446}
\BIBentrySTDinterwordspacing

\bibitem{barry2022}
\BIBentryALTinterwordspacing
P.~S. Barry, A.~Anderson, B.~Benson, J.~E. Carlstrom, T.~Cecil, C.~Chang,
  M.~Dobbs, M.~Hollister, K.~S. Karkare, G.~K. Keating, D.~Marrone, J.~McMahon,
  J.~Montgomery, Z.~Pan, G.~Robson, M.~Rouble, E.~Shirokoff, and G.~Smecher,
  ``{Design of the~{SPT}-{SLIM} Focal Plane: A Spectroscopic Imaging Array for
  the South Pole Telescope},'' \emph{Journal of Low Temperature Physics}, vol.
  209, no. 5-6, pp. 879--888, sep 2022. [Online]. Available:
  \url{https://doi.org/10.1007}
\BIBentrySTDinterwordspacing

\bibitem{endo2012}
\BIBentryALTinterwordspacing
A.~Endo, J.~J.~A. Baselmans, P.~P. van~der Werf, B.~Knoors, S.~M.~H.
  Javadzadeh, S.~J.~C. Yates, D.~J. Thoen, L.~Ferrari, A.~M. Baryshev, Y.~J.~Y.
  Lankwarden, P.~J. de~Visser, R.~M.~J. Janssen, and T.~M. Klapwijk,
  ``{Development of DESHIMA: a redshift machine based on a superconducting
  on-chip filterbank},'' in \emph{Millimeter, Submillimeter, and Far-Infrared
  Detectors and Instrumentation for Astronomy VI}, W.~S. Holland, Ed., vol.
  8452, International Society for Optics and Photonics.\hskip 1em plus 0.5em
  minus 0.4em\relax SPIE, 2012, p. 84520X. [Online]. Available:
  \url{https://doi.org/10.1117/12.925637}
\BIBentrySTDinterwordspacing

\bibitem{shirokoff2012}
\BIBentryALTinterwordspacing
E.~Shirokoff, P.~S. Barry, C.~M. Bradford, G.~Chattopadhyay, P.~Day, S.~Doyle,
  S.~Hailey-Dunsheath, M.~I. Hollister, A.~Kov{\'a}cs, C.~McKenney, H.~G.
  Leduc, N.~Llombart, D.~P. Marrone, P.~Mauskopf, R.~O'Brient, S.~Padin,
  T.~Reck, L.~J. Swenson, and J.~Zmuidzinas, ``{MKID development for SuperSpec:
  an on-chip, mm-wave, filter-bank spectrometer},'' in \emph{Millimeter,
  Submillimeter, and Far-Infrared Detectors and Instrumentation for Astronomy
  VI}, W.~S. Holland, Ed., vol. 8452, International Society for Optics and
  Photonics.\hskip 1em plus 0.5em minus 0.4em\relax SPIE, 2012, p. 84520R.
  [Online]. Available: \url{https://doi.org/10.1117/12.927070}
\BIBentrySTDinterwordspacing

\bibitem{wilson2020}
\BIBentryALTinterwordspacing
G.~W. Wilson, S.~Abi-Saad, P.~Ade, I.~Aretxaga, and e.~a. Jason~Austermann,
  ``{The TolTEC camera: an overview of the instrument and in-lab testing
  results},'' in \emph{Millimeter, Submillimeter, and Far-Infrared Detectors
  and Instrumentation for Astronomy X}, J.~Zmuidzinas and J.-R. Gao, Eds., vol.
  11453, International Society for Optics and Photonics.\hskip 1em plus 0.5em
  minus 0.4em\relax SPIE, 2020, p. 1145302. [Online]. Available:
  \url{https://doi.org/10.1117/12.2562331}
\BIBentrySTDinterwordspacing

\bibitem{brien2018}
\BIBentryALTinterwordspacing
T.~L.~R. Brien, P.~A.~R. Ade, P.~S. Barry, E.~Castillo-Dom{\`i}nguez,
  D.~Ferrusca, T.~Gascard, V.~G{\'o}mez, P.~C. Hargrave, A.~L. Hornsby,
  D.~Hughes, E.~Pascale, J.~D.~A. Parrianen, A.~Perez, S.~Rowe, C.~Tucker,
  S.~V. Gonz{\'a}lez, and S.~M. Doyle, ``{MUSCAT: the Mexico-UK Sub-Millimetre
  Camera for AsTronomy},'' in \emph{Millimeter, Submillimeter, and Far-Infrared
  Detectors and Instrumentation for Astronomy IX}, J.~Zmuidzinas and J.-R. Gao,
  Eds., vol. 10708, International Society for Optics and Photonics.\hskip 1em
  plus 0.5em minus 0.4em\relax SPIE, 2018, p. 107080M. [Online]. Available:
  \url{https://doi.org/10.1117/12.2313697}
\BIBentrySTDinterwordspacing

\bibitem{calvo2016}
\BIBentryALTinterwordspacing
M.~Calvo, A.~Beno{\^i}t, A.~Catalano, J.~Goupy, A.~Monfardini, N.~Ponthieu,
  E.~Barria, and G.~e.~a. Bres, ``The nika2 instrument, a dual-band kilopixel
  kid array for millimetric astronomy,'' \emph{Journal of Low Temperature
  Physics}, vol. 184, no.~3, pp. 816--823, Aug 2016. [Online]. Available:
  \url{https://doi.org/10.1007/s10909-016-1582-0}
\BIBentrySTDinterwordspacing

\bibitem{thesisGao}
J.~Gao, ``The physics of superconducting microwave resonators,'' Ph.D. diss.,
  Pasadena CA, 2008.

\bibitem{thesispete}
B.~Pete, ``On the development of superspec; a fully integrated on-chip
  spectrometer for far-infrared astronomy,'' Ph.D. diss., Cardiff, United
  Kindom, 2014.

\bibitem{visserprl}
\BIBentryALTinterwordspacing
P.~J. de~Visser, J.~J.~A. Baselmans, P.~Diener, S.~J.~C. Yates, A.~Endo, and
  T.~M. Klapwijk, ``Number fluctuations of sparse quasiparticles in a
  superconductor,'' \emph{Phys. Rev. Lett.}, vol. 106, p. 167004, Apr 2011.
  [Online]. Available:
  \url{https://link.aps.org/doi/10.1103/PhysRevLett.106.167004}
\BIBentrySTDinterwordspacing

\bibitem{kaplan1976}
\BIBentryALTinterwordspacing
S.~B. Kaplan, C.~C. Chi, D.~N. Langenberg, J.~J. Chang, S.~Jafarey, and D.~J.
  Scalapino, ``Quasiparticle and phonon lifetimes in superconductors,''
  \emph{Phys. Rev. B}, vol.~14, pp. 4854--4873, Dec 1976. [Online]. Available:
  \url{https://link.aps.org/doi/10.1103/PhysRevB.14.4854}
\BIBentrySTDinterwordspacing

\bibitem{catelani2011}
\BIBentryALTinterwordspacing
G.~Catelani, J.~Koch, L.~Frunzio, R.~J. Schoelkopf, M.~H. Devoret, and L.~I.
  Glazman, ``Quasiparticle relaxation of superconducting qubits in the presence
  of flux,'' \emph{Phys. Rev. Lett.}, vol. 106, p. 077002, Feb 2011. [Online].
  Available: \url{https://link.aps.org/doi/10.1103/PhysRevLett.106.077002}
\BIBentrySTDinterwordspacing

\bibitem{martinis2009}
\BIBentryALTinterwordspacing
J.~M. Martinis, M.~Ansmann, and J.~Aumentado, ``Energy decay in superconducting
  josephson-junction qubits from nonequilibrium quasiparticle excitations,''
  \emph{Phys. Rev. Lett.}, vol. 103, p. 097002, Aug 2009. [Online]. Available:
  \url{https://link.aps.org/doi/10.1103/PhysRevLett.103.097002}
\BIBentrySTDinterwordspacing

\bibitem{visser2012}
\BIBentryALTinterwordspacing
P.~J. de~Visser, J.~J.~A. Baselmans, S.~J.~C. Yates, P.~Diener, A.~Endo, and
  T.~M. Klapwijk, ``Microwave-induced excess quasiparticles in superconducting
  resonators measured through correlated conductivity fluctuations,''
  \emph{Applied Physics Letters}, vol. 100, no.~16, p. 162601, 2012. [Online].
  Available: \url{https://doi.org/10.1063/1.4704151}
\BIBentrySTDinterwordspacing

\end{thebibliography}

\end{document}